\documentclass[onecolumn,showpacs]{revtex4}

\usepackage{graphicx}% Include figure files
\usepackage{dcolumn}% Align table columns on decimal point
\usepackage{bm}% bold math

\input epsf

\begin{document}

\title{\Large Effect of Dynamical Cosmological Constant in presence of Modified Chaplygin Gas for Accelerating Universe}

\author{\bf Writambhara Chakraborty$^1$\footnote{writam1@yahoo.co.in}
and Ujjal Debnath$^2$\footnote{ujjaldebnath@yahoo.com} }

\affiliation{$^1$Department of Mathematics, Heritage Institute of
Technology, Anandapur, Kolkata-700 107, India.\\
$^2$Department of Mathematics, Bengal Engineering and Science
University, Shibpur, Howrah-711 103, India. }

\date{\today}

\begin{abstract}
In this paper we have considered the Universe to be filled with
Modified Gas and the Cosmological Constant $\Lambda$ to be
time-dependent with or without the Gravitational Constant $G$ to
be time-dependent. We have considered various phenomenological
models for $\Lambda$ , viz., $\Lambda\propto\rho,
\Lambda\propto\frac{\dot{a}^{2}}{a^{2}}$  and
$\Lambda\propto\frac{\ddot{a}}{a}$. Using these models it is
possible to show the accelerated expansion of the Universe at the
present epoch. Also we have shown the natures of $G$ and
$\Lambda$ over the total age of the Universe. Using the
statefinder parameters we have shown the diagramamtical
representation of the evolution of the Universe starting from
radiation era to $\Lambda$CDM model.
\end{abstract}

\pacs{}

\maketitle

\section{\normalsize\bf{Introduction}}

There are two parameters, the cosmological constant $\Lambda$ and
the gravitational constant $G$, present in Einstein's field
equations. The Newtonian constant of gravitation $G$ plays the
role if a coupling constant between geometry and matter in the
Einstein's field equations. In an evolving Universe, it appears
natural to look at this ``constant'' as a function of time.
Numerous suggestions based on different arguments have been
proposed in the past few decades in which $G$ varies with time
[1]. Dirac [2] proposed a theory with variable $G$ motivated by
the occurrence of large numbers discovered by Weyl, Eddington and
Dirac himself.\\

It is widely believed that the value of $\Lambda$ was large during
the early stages of evolution and strongly influenced its
expansion, whereas its present value is incredibly small [3].
Several authors [4] have advocated a variable $\Lambda$ in the
framework of Einstein's theory to account for this fact.
$\Lambda$ as a function of time has also been considered in
various variable $G$ theories in different contexts [5]. For
these variations, the energy-momentum tensor of matter leaves
the form of the Einstein's field equations unchanged.\\

In attempt to modify the General Theory of Relativity, Al-Rawaf
and Taha [6] related the cosmological constant to the Ricci
Scalar $\cal R$. This is written as a built-in-cosmological
constant, i.e., $\Lambda\propto\cal R$. Since the Ricci Scalar
contains a term of the form $\frac{\ddot{a}}{a}$, one adopts this
variation for $\Lambda$. We parameterized this as
$\Lambda\propto\frac{\ddot{a}}{a}$. Similarly, we have chosen
another two forms for $\Lambda : \Lambda\propto\rho$ and
$\Lambda\propto\frac{\dot{a}^{2}}{a^{2}}$; where $\rho$ is the
energy density. \\

Recent observations of the luminosity of type Ia Supernovae
indicate [7, 8] an accelerated expansion of the Universe and lead
to the search for a new type of matter which violates the strong
energy condition i.e., $\rho+3p<0$. The matter consent
responsible for such a condition to be satisfied at a certain
stage of evolution of the universe is referred to as a {\it dark
energy}. There are different candidates to play the role of the
dark energy. The type of dark energy represented by a scalar
field is often called Quintessence. The simplest candidate for
dark energy is Cosmological Constant $\Lambda$. In particular one
can try another type of dark energy, the so-called Chaplygin gas
which obeys an  equation of state like [9] $p=-B/\rho, (B>0)$,
where $p$ and $\rho$ are respectively the pressure and energy
density. Subsequently the above equation was modified to the form
[10] $p=-B/\rho^{\alpha}, 0\le \alpha \le 1$. There are some
works on Modified Chapligin Gas obeying equation of state [11, 12]

\begin{equation}
p=A\rho-B/\rho^{\alpha}, (A>0)
\end{equation}

In this work we have considered a cosmological model for the
cosmological constant of the forms :
$\Lambda\propto\frac{\ddot{a}}{a}$,
$\Lambda\propto\frac{\dot{a}^{2}}{a^{2}}$ and
$\Lambda\propto\rho$ in presence of Modified Chaplygin Gas, with
or without the variation of Gravitational Constant $G$.\\

In 2003, V. Sahni et al [13] introduced a pair of parameters
$\{r,s\}$, called {\it statefinder} parameters. The trajectories
in the $\{r,s\}$ plane corresponding to different cosmological
models demonstrate qualitatively different behaviour. The above
statefinder diagnostic pair has the following form:

\begin{equation}
r=\frac{\dddot{a}}{aH^{3}}~~~~\text{and}~~~~s=\frac{r-1}{3\left(q-\frac{1}{2}\right)}
\end{equation}

where $H\left(=\frac{\ddot{a}}{a}\right)$ and
$q~\left(=-\frac{a\ddot{a}}{\dot{a}^{2}}\right)$ are the Hubble
parameter and the deceleration parameter respectively. The new
feature of the statefinder is that it involves the third
derivative of the cosmological radius. These parameters are
dimensionless and allow us to characterize the properties of dark
energy. Trajectories in the $\{r,s\}$ plane corresponding to
different cosmological models, for example $\Lambda$CDM model
diagrams correspond to the fixed point $s=0,~r=1$.\\

The paper is organized as follows: Section II deals with Einstein
field equations in presence of dynamic cosmological constant
$\Lambda$. In sections III and IV we have considered different
$\Lambda$-dependent models without and with variable
gravitational constant $G$ respectively and shown various stages
of the evolution of the Universe for these models using
statefinder parameters. We have taken some particular values of
the proportionality constants for the graphical representations.
We have discussed this particularization of the constants and
their physical consequences
in section V.\\

\section{\normalsize\bf{Einstein Field Equations and Dynamic Cosmological Constant}}

We consider the spherically symmetric FRW metric,

\begin{equation}
ds^{2}=dt^{2}-a^{2}(t)\left[\frac{dr^{2}}{1-kr^{2}}+r^{2}(d\theta^{2}+sin^{2}\theta
d\phi^{2})\right]
\end{equation}

where $a(t)$ is the scale factor and $k$ is the curvature scalar
with values  0, 1 and $-$1 for respectively  flat, closed and open
models of the Universe. The Einstein field equations for a
spatially flat Universe (i.e., taking $k=0$) with a time-dependent
cosmological constant $\Lambda(t)$ are given by (choosing $c=1$),

\begin{equation}
3\frac{\dot{a}^{2}}{a^{2}}=8\pi G \rho+\Lambda(t)
\end{equation}
and
\begin{equation}
2\frac{\ddot{a}}{a}+\frac{\dot{a}^{2}}{a^{2}}=-8\pi G p+\Lambda(t)
\end{equation}

where $\rho$ and $p$ are the energy density and isotropic presure
respectively.\\

Let us choose modified chaplygin gas with equation of state given
by equation (1).\\

Here, we consider the phenomenological models for $\Lambda(t)$ of
the forms $\Lambda\propto\rho$,
$\Lambda\propto\frac{\dot{a}^{2}}{a^{2}}$ and
$\Lambda\propto\frac{\ddot{a}}{a}$.\\
First we will consider $G$ to be constant and try to find out the
solutions for density $\rho$ and the scale factor $a(t)$ and hence
study the cosmological models in terms of the statefinder
parameters $r$, $s$. Secondly we will consider $G$ to be variable
as well and study the various phases of the Universe represented
by the models.\\

\section{\normalsize\bf{Models keeping $G$ constant and $\Lambda$ variable}}

Taking $G$ to be constant and $\Lambda$ to be time dependent, the
energy conservation equation is,
\begin{equation} \dot{\rho}+3\frac{\dot{a}}{a}(\rho+p)=-\frac{\dot{\Lambda}}{8\pi G}
\end{equation}\\

\subsection{\normalsize\bf{Model with $\Lambda\propto\rho$}}

Here we consider \begin{equation} \Lambda=\beta_{1}~\rho
\end{equation}
where $\beta_{1}$ is a constant.\\

Equation (7) together with equations (1) and (6) yield the
solution for $\rho$ to be,
\begin{equation}\rho=\left(\frac{B}{1+A}+\frac{C}{a^{\frac{24\pi G(1+A)(1+\alpha)}{8\pi G+\beta_{1}}}}\right)^{\frac{1}{1+\alpha}}\end{equation}
where $C$ is an arbitrary constant.\\

Substituting equation (7) and (8) in equation (4), we get the
solution for the scale factor $a(t)$ as,
\begin{equation}
a^{f_{1}f_{2}}\sqrt{8 \pi G+\beta_{1}}~~_{2}F_{1}
[f_{2},f_{2},1+f_{2},-\frac{a^{f_{1} B}}{C
(1+A)}]=4\sqrt{3}(1+A)G\pi C^{f_{2}}~t
\end{equation}
where  $f_{1}=\frac{24(1+A)(1+\alpha)\pi G }{8 \pi G +\beta_{1}}$
and  $f_{2}=\frac{1}{2(1+\alpha)}$. Hence, for small values of
$a(t)$, we have,  $\rho\simeq\left(\frac{C}{a^{\frac{24 \pi
G(1+A)(1+\alpha}{8 \pi
G+\beta_{1}}}}\right)^{\frac{1}{1+\alpha}}$ which is very large
and the equation of state (1) reduces to $p\simeq A \rho$. Again
for large values of $a(t)$, we get
$\rho\simeq\left(\frac{B}{1+A}\right)^\frac{1}{1+\alpha}$ and
$p\simeq-\left(\frac{B}{1+A}\right)^\frac{1}{1+\alpha}$ , i.e.,
$p\simeq-\rho$ which coincides with the result obtained for MCG
with $\beta_{1}=0$ [12].\\

\begin{figure}
\includegraphics[height=1.7in]{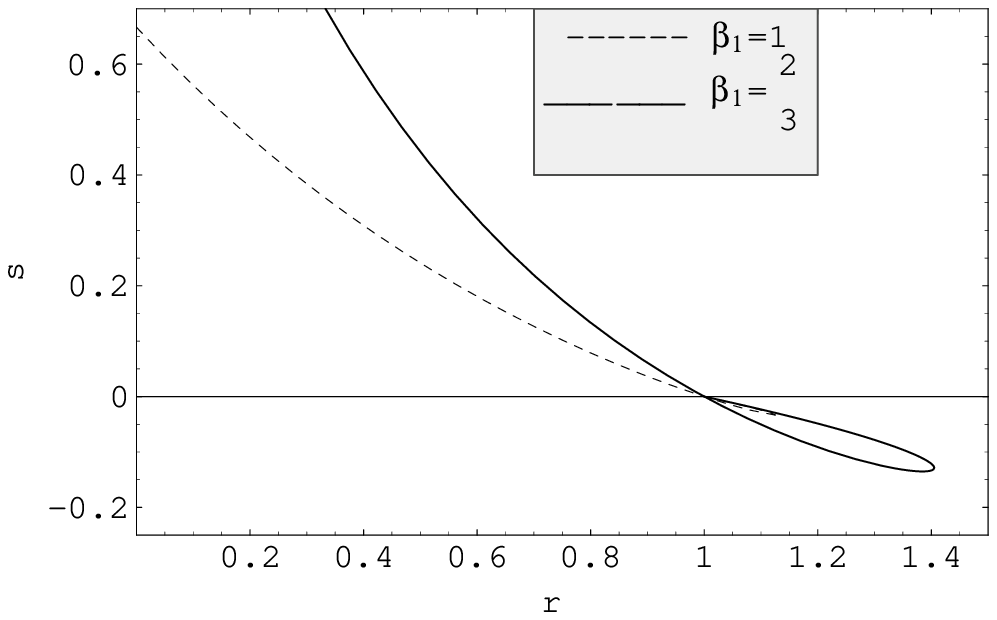}

Fig.1 \vspace{5mm}

\vspace{5mm} Fig. 1 shows the variation of $s$
 against $r$ for different values of $\beta_{1}=1,~\frac{2}{3}$ respectively and for
 $\alpha$ = ~1, $A=1/3$ and $8~\pi~ G~ $=~1. \hspace{14cm} \vspace{4mm}

\end{figure}

Using equations (2), (4) and (6) we get,
\begin{equation}
r=1+\frac{36 \pi G(1+y)[8 \pi G
\{A(1+\alpha)-y\alpha\}-\beta_{1}]}{(8 \pi
G+\beta_{1})^{2}}~~~\text{and}~~~s=\frac{8\pi G (1+y)[8\pi G
\{A(1+\alpha)-y\alpha\}-\beta_{1}] }{(8\pi G +\beta_{1})(8\pi G
y-\beta_{1})}
\end{equation}
where $y=\frac{p}{\rho}$ which can be further reduced to a single
relation between $r$ and $s$. Now
$q=-\frac{\ddot{a}}{aH^{2}}=\frac{8\pi G (1+3y)-2\beta_{1}}{2(8\pi
G+\beta_{1})}$. Therefore for acceleration
$q<0~\Rightarrow~y<\frac{\beta_{1}}{12\pi G}-\frac{1}{3}$. Also
for the present epoch
$q=-\frac{1}{2}~\Rightarrow~y=\frac{1}{3}(\frac{3\beta_{1}}{8\pi
G}-2)$. If we assume that the present Universe is dust filled, we
have $y=0$, i.e., $\beta_{1}=\frac{16\pi G}{3}$. Taking $8\pi G=1$
we get the best fit value to be $\beta_{1}=\frac{2}{3}$, which
gives $r=1$ ( choosing $A=\frac{1}{3},~\alpha=1$ ) for the
present Universe. That means the dark energy responsible for the
the present acceleration is nothing but $\Lambda$. Also $A=1,~
\alpha=1$ and $\beta_{1}=\frac{2}{3}$ give $r=2.16$ for the
present time. For this case $\beta_{1}>3$ gives non-feasible
solutions in the sense that the present values of $y$, i.e.,
$\frac{p}{\rho}$ becomes too large. For $\beta_{1}=1$, we get the
present value of $y$ to be $\frac{1}{3}$, but again $r=1$. In
either of the above cases we get accelerating expansion of the
Universe. These can be represented diagrammatically in the
{$r,s$} plane. This is shown in figure 1 (taking $A=\frac{1}{3},
\alpha=1, \beta_{1}=1, \frac{2}{3}, 8\pi G=1$ and $A=1, \alpha=1,
\beta_{1}=1,\frac{2}{3}, 8\pi G=1$). Figure 1 represents the
evolution of the Universe starting from radiaton era to
$\Lambda$CDM model.
Here we get a discontinuity at $\beta_{1}=-8\pi G$. \\

Again for this model
\begin{equation}
\Lambda=\beta_{1}\left(\frac{B}{1+A}+\frac{C}{a^{\frac{24\pi
G(1+A)(1+\alpha)}{8\pi G+\beta_{1}}}}\right)^{\frac{1}{1+\alpha}}
\end{equation}
Variation of $\Lambda(t)$ against $a(t)$ is shown in figure 2 for
different choices of $\beta_{1}$, which represents that
regardless the values of $\beta_{2}$, $\Lambda(t)$, i.e., the
effect of the cosmological constant decreases with time.\\

\begin{figure}
\includegraphics[height=1.7in]{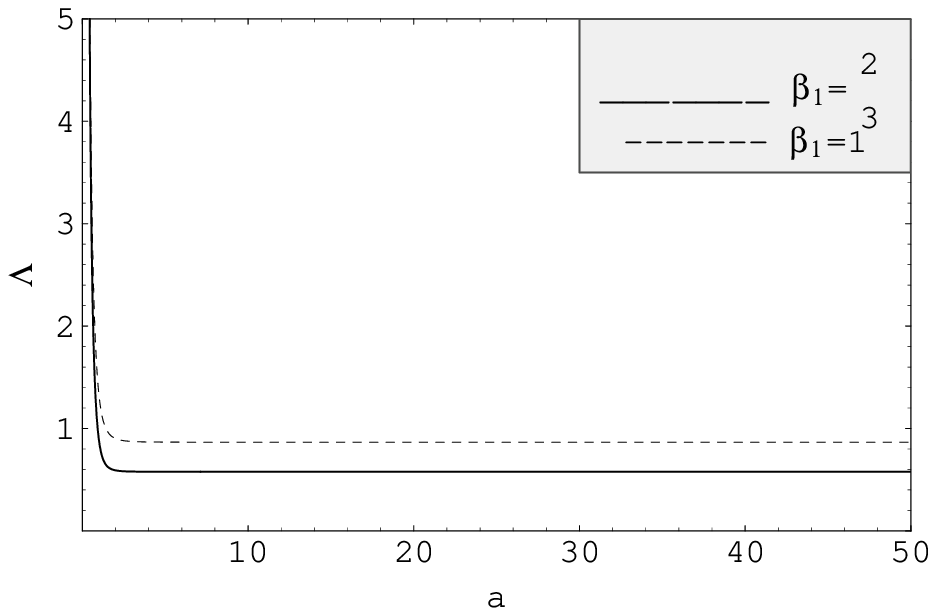}
\\
\vspace{1mm} Fig.2 \vspace{5mm}

\vspace{5mm} Fig. 2 shows the variation of $\Lambda$
 against $a(t)$ for different values of $\beta_{1}=~1,~\frac{2}{3}$ respectively and for
 $\alpha$ = ~1, $A=1/3$, $8~\pi~G$ =~1, $B$ =~1, $C$ =~1 .\hspace{14cm} \vspace{4mm}

\end{figure}

\subsection{\normalsize\bf{Model with $\Lambda\propto H^{2}$}}

Choosing \begin{equation} \Lambda(t)=\beta_{2}H^{2} \end{equation}
where $\beta_{2}$ is a constant and proceeding as above, we
obtain the solutions for $\rho, a(t), \Lambda$ as,
\begin{equation}\rho=\left(\frac{B}{1+A}+\frac{C}{a^{(3-\beta_{2})(1+A)(1+\alpha)}}\right)^{\frac{1}{1+\alpha}}\end{equation}
\begin{equation}
a^{f_{1}f_{2}}~_{2}F_{1} [f_{2},f_{2},1+f_{2},-\frac{a^{f_{1}
B}}{C (1+A)}]=\sqrt{2 \pi G }\sqrt{3-\beta_{2}}~(1+A) C^{f_{2}}~t
\end{equation}
where  $f_{1}=(3-\beta_{2})(1+A)(1+\alpha)$ and
$f_{2}=\frac{1}{2(1+\alpha)}$
\begin{equation}
\Lambda=\frac{8 \pi G
\beta_{2}}{3-\beta_{2}}\left(\frac{B}{1+A}+\frac{C}{a^{(3-\beta_{2})(1+A)(1+\alpha)}}\right)^{\frac{1}{1+\alpha}}
\end{equation}
Here for $\beta_{2} < 3$ we can check the consistency of the
result by showing $p\simeq A\rho$ at small values of $a(t)$ and
$p=-\rho$ for large values of $a(t)$. But if we take $\beta_{2} >
3$ we get opposite results which contradict our previous notions
of the nature of the equation of state (1). Again for $\beta_{2} =
3$, we get only $\Lambda$CDM point ,i.e., we get a discontinuity.
Therefore, we restrict our choice for $\beta_{2}$
in this case to be $\beta_{2} < 3 $.\\

\begin{figure}
\includegraphics[height=1.7in]{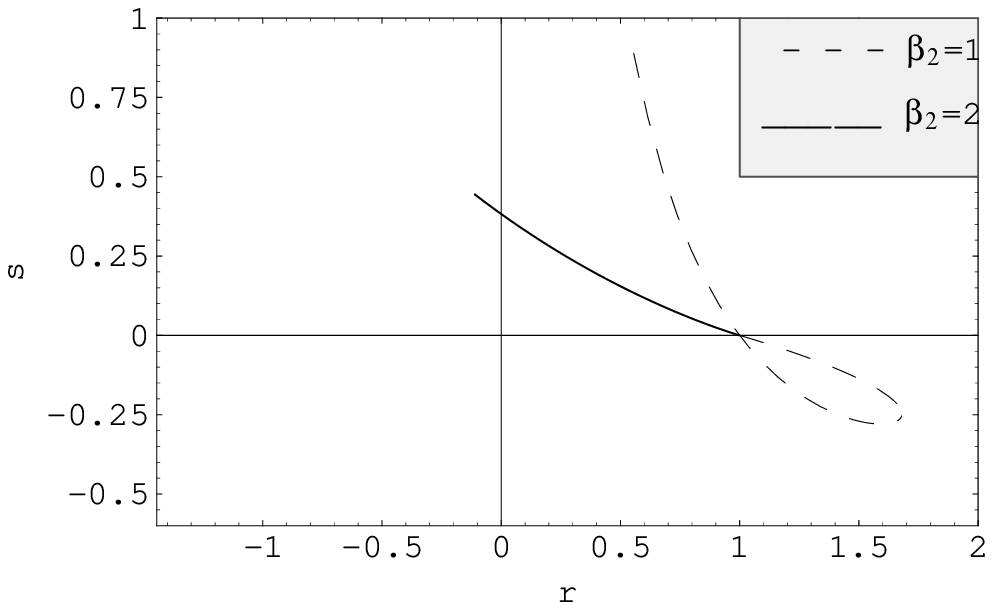}
\includegraphics[height=1.7in]{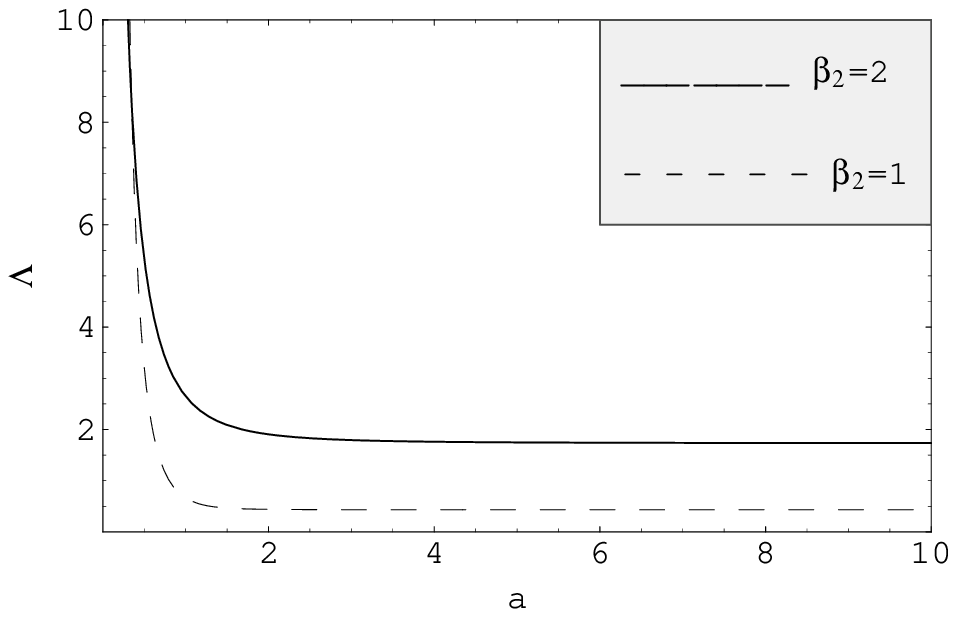}\\
\vspace{1mm}
Fig.3~~~~~~~~~~~~~~~~~~~~~~~~~~~~~~~~~~~~~~~~~~~~~~~~~~~~~Fig.4\\
\vspace{5mm}

\vspace{5mm} Fig. 3 shows the variation of $s$
 against $r$ for different values of $\beta_{2}=1,~2$ and for
 $\alpha$ = ~1, $A=1/3$, $8~\pi~G~=~1$. Fig. 4 shows the variation of $\Lambda$ against $a(t)$ for
 different values of $\beta_{2}=1,~2$ and for
 $\alpha$ = ~1, $A=1/3$, $8~\pi~G~=~1$. \hspace{14cm} \vspace{4mm}

\end{figure}

Computing the state-finder parameters given by equation (2), we
get the equations for $r$ and $s$ to be,
\begin{equation}
r=1+\frac{(3-\beta_{2})(1+y)[\{A(1+\alpha)-y\alpha\}(3-\beta_{2})-\beta_{2}]}{2}~~~\text{and}~~~s=\frac{(3-\beta_{2})(1+y)][\{A(1+\alpha)-y\alpha\}(3-\beta_{2})-\beta_{2}]}{3\{(3-\beta_{2})y-\beta_{2}\}}
\end{equation}
(where $y=\frac{p}{\rho}$), which can still be resolved into a
single relation and can be plotted in the ${r,s}$ plane. Here
$q=\frac{1}{2}[(3-\beta_{2})y-(\beta_{2}-1)]$. Hence the Universe
will accelerate if
$q<0~\Rightarrow~y<\frac{\beta_{2}-1}{3-\beta_{2}}$. Again for
the present Universe
$q=-\frac{1}{2}~\Rightarrow~y=\frac{\beta_{2}-2}{3-\beta_{2}}$.
Assuming the present Universe to be dust dominated, i.e., $y=0$
we get the best fit value for $\beta_{2}$ to be $2$. Taking
$A=\frac{1}{3},\alpha=1,8\pi G=1,\beta_{2}=2$ and $y=0$( i.e.
dust dominated present Universe ) we get the present value to be
$r=1/3$, also the same values with $\beta_{2}=1$ gives the
present values to be $y=-\frac{1}{2},~r=\frac{5}{3}$ .This is
shown in figure 3 ($A=\frac{1}{3}, \alpha=1,\beta_{2}=1$ and $2,
8 \pi G=1$), which explains the evolution of the Universe from
radiation era to $\Lambda$CDM model. Again  variation of
$\Lambda$ against time is shown in figure 4, where we can see that
$\Lambda$ decreases with time for whatever the value of $\beta_{2}$ be.\\

\subsection{\normalsize\bf{Model with $\Lambda\propto \frac{\ddot{a}}{a}$}}

Taking \begin{equation} \Lambda=\beta_{3}\frac{\ddot{a}}{a}
\end{equation} (where $\beta_{3}$ is a constant), and proceeding
as above we get a relation for $\rho$ as,
\begin{equation}
\rho^{(\frac{2}{1+A}-\beta_{3})}\left(1+A-\frac{B}{\rho^{\alpha+1}}\right)^{(\frac{2}{(1+A)(1+\alpha)}-\beta_{3})}=\frac{C}{a^{2(3-\beta_{3})}}
\end{equation}Unlike the previous two cases here we get a far more
restricted solution. Here the only choice of $\beta_{3}$ for
which we get the feasible solution satisfying  $p\simeq A\rho$ for
small values of $a(t)$ and $p\simeq -\rho$ for large values of
$a(t)$ is
\begin{equation} \beta_{3}<\frac{2}{(1+A)(1+\alpha)}~~~~~~\text
{or}~~~~~~ \beta_{3}>3
\end{equation}
Again since $q=-\frac{\ddot{a}}{a
H^{2}}=-\frac{\Lambda}{\beta_{3} H^{2}}=\frac{4\pi
G(\rho+3p)}{(3-\beta_{3})H^{2}}=\frac{4\pi G
(\rho+3p)}{(3-\beta_{3})H^{2}}$, $\beta_{3}>3$ implies $q<0$
without even violating the energy-condition $\rho+3p\geq 0$.
Although $\beta_{3}<\frac{2}{(1+A)(1+\alpha)}$ causes the
acceleration of the Universe violating the energy-condition.
Taking $q=-\frac{1}{2}$ for the present epoch, we obtain
$y=\frac{(\beta_{3}-4)}{(6-\beta_{3})}$. Hence the present epoch
is dust filled if $\beta{3}=4$ and thus giving the present value
of $r$ to be $-\frac{1}{7}$ for $A=\frac{1}{3},~\alpha=1,~8\pi
G=1$.On using relation (19), $\rho$ and therefore $a,~\Lambda$
cannot be expressed in an open form. We can rather derive a
solution for $\Lambda$ in terms of $p,~ \rho$ as,
\begin{equation}
\Lambda=\frac{4 \pi G \beta_{3}}{\beta_{3}-3}(\rho+3p)
\end{equation}
Using equations (2)we get the statefinder parameters as,
\begin{equation}
r=1-\frac{(1+y)(\beta_{3}-3)[\beta_{3}+(\beta_{3}+6)x]}{(\beta_{3}+\beta_{3}x-2)(\beta_{3}+\beta_{3}y-2)}~~~\text{and}~~~s=\frac{2(1+y)(\beta_{3}-3)[\beta_{3}+(\beta_{3}+6)x]}{[\beta_{3}+(\beta_{3}+6)y][\beta_{3}+\beta_{3}x-2]
}
\end{equation}
where $y=\frac{p}{\rho}$ and $x=\frac{\partial p}{\partial
\rho}$, i.e., $x=A(1+\alpha)-y\alpha$ [from equation (1)].\\

\begin{figure}
\includegraphics[height=1.7in]{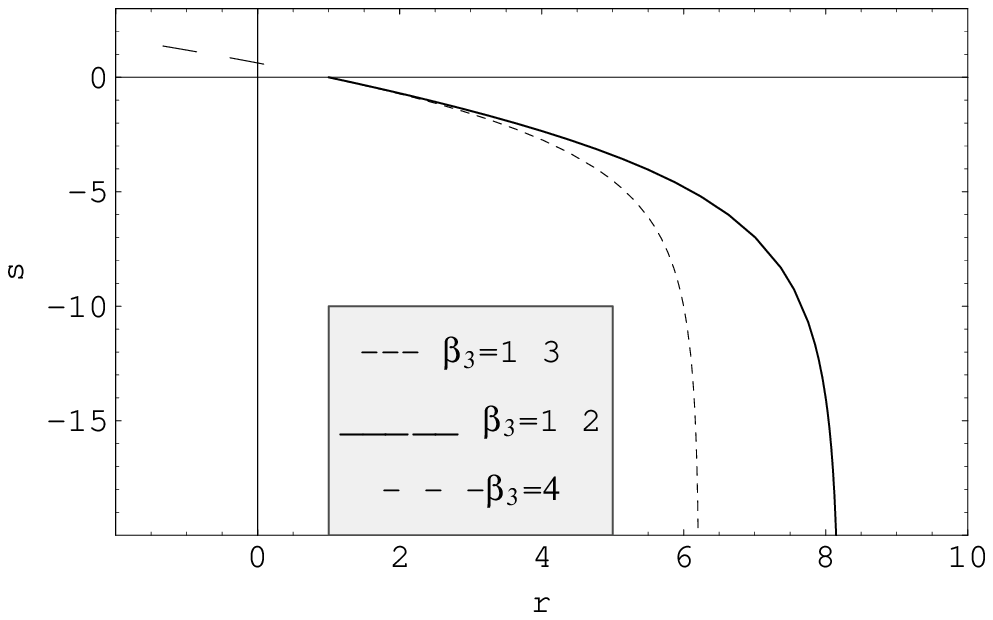}\\
\vspace{1mm}
Fig.5\\
\vspace{5mm}

\vspace{5mm} Fig. 5 shows the variation of $s$
 against $r$ for different values of $A=\frac{1}{3}, \alpha=1, 8\pi G=1, \beta_{3}=\frac{1}{2},4$ and
$\frac{1}{3}$\hspace{14cm} \vspace{4mm}
 \end{figure}

Eliminating $y$ between the equations (21), we get a single
relation of $r$ and $s$, which can be represented diagrammatically
in the ${r,s}$ plane (figure 5). Here we have taken
$A=\frac{1}{3}, \alpha=1, 8\pi G=1, \beta_{3}=\frac{1}{2},4$ and
$\frac{1}{3}$, combining two cases. Taking
$\beta_{3}=\frac{1}{2},\frac{1}{3}$ we can explain the evolution
of the Universe starting from $\frac{p}{\rho}=-\frac{1}{3}$ to
$\Lambda$CDM model and $\beta_{3}=4$ explains the evolution of the
Universe starting from radiation era to $y=-\frac{1}{3}$, as seen
from the expression for $q$. Considering the present epoch to be
dust-dominated, the present value of $r$ is given for
$\beta_{3}=4$ to be $-\frac{1}{7}$. As follows, the former two
cases cannot give the present value of $r$, as $y=0>-\frac{1}{3}$
for the present epoch. Here we have an infinite discontinuity at
$\frac{p}{\rho}=-\frac{1}{3}$, i.e.,
when $\rho+3p=0$. Also since we do not get a closed from of $\rho$ here, it is difficult to plot $\Lambda$ against the scale factor $a(t)$ .\\

\section{\normalsize\bf{Models with $\Lambda$ and $G$ both variable}}

Now we consider $G$ as well as $\Lambda$ to be variable. With
this the equations (3), (4), (5) yield the conservation laws as,
\begin{equation}
\dot{\rho}+3H(\rho+p)=0
\end{equation}
and
\begin{equation} \dot{\Lambda}+ 8 \pi \dot{G} \rho=0
\end{equation}
Now we study the various phases of the Universe represented by
these models.\\

Equation (22) together with equation (1) yield the solution for
$\rho$ as,
\begin{equation}\rho=\left(\frac{B}{1+A}+\frac{C}{a^{3(1+A)(1+\alpha)}}\right)^{\frac{1}{1+\alpha}}\end{equation}
where $C$ is an arbitrary constant. This result is consistent
with the results already obtained [12].\\

\subsection{\normalsize\bf{Model with $\Lambda\propto \rho$}}

Here we consider \begin{equation} \Lambda= \gamma_{1}~ \rho
\end{equation}
where $\gamma_{1}$ is a constant.\\

Equation (22), (23) and (25) give,
\begin{equation}
G=C_{1}-\frac{\gamma_{1}}{8\pi}\log\rho
\end{equation} where $C_{1}$ is a constant and $\rho$ is given by
equation (24).\\

Using equations (2), (4), (23) and (26), we get,
\begin{eqnarray}
\begin{array} {ccc}
G=C_{1}+\frac{\gamma_{1}(1+\alpha)}{8\pi}\log(\frac{B}{A-y})\\\\\
r=1+\frac{9(1+y)[8\pi G
\{A(1+\alpha)-y\alpha\}-\gamma_{1}(1+y)]}{2(8\pi
G+\gamma_{1})}\\\\\
s=\frac{(1+y)[8\pi G
\{A(1+\alpha)-y\alpha\}-\gamma_{1}(1+y)]}{(8\pi G y-\gamma_{1})}\\\\
\end{array}
\end{eqnarray}
where $y=\frac{p}{\rho}$.\\

\begin{figure}
\includegraphics[height=1.7in]{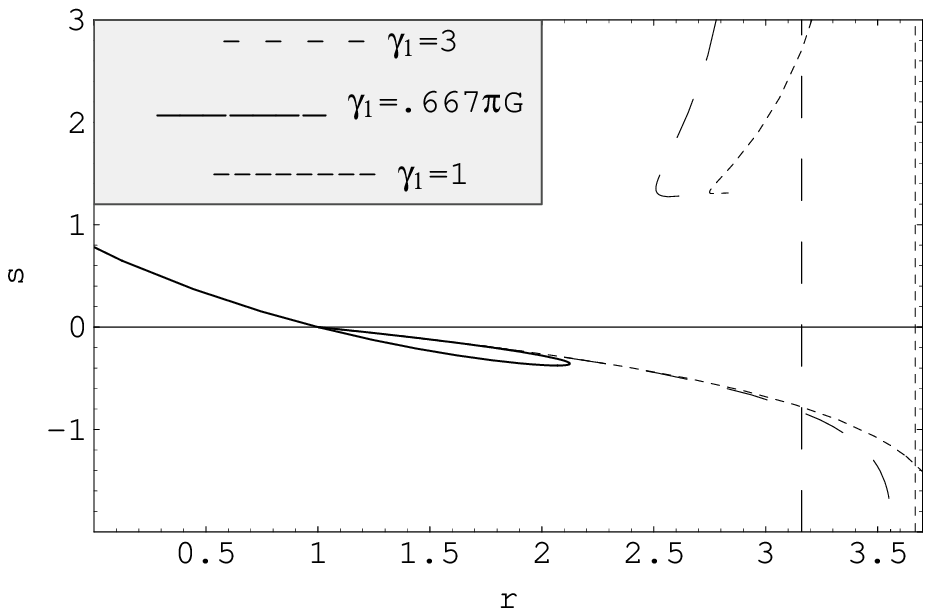}

Fig.6 \vspace{5mm}

\vspace{5mm} Fig. 6 shows the variation of $s$
 against $r$ for different values of $\gamma_{1}=1, 3$ and $3.5$ and
$A=\frac{1}{3}, \alpha=1, B=1, C_{1}=1$. \hspace{14cm}
\vspace{4mm}

\end{figure}

Equation (27) cannot be resolved to get a single relation between
$r$ and $s$, rather we obtain a parametric relation between the
same with $y=\frac{p}{\rho}$ as the parameter. This can be
represented diagrammatically in the ${r,s}$ plane, which is shown
in figure 6 taking $\gamma_{1}=1, 3$ and $3.5$ and $A=\frac{1}{3},
\alpha=1, B=1, C_{1}=1$. Now $q=\frac{4\pi G
(1+3y)-\gamma_{1}}{8\pi G+\gamma_{1}}$. Taking into account that
$q=-\frac{1}{2}$ for the present epoch, we get
$y=\frac{\gamma_{1}}{8\pi G}-\frac{2}{3}$. Therefore, for the
present dust-dominated era $y=0$ and $\gamma_{1}=\frac{16 \pi
G}{3}$. hence for the This models represents the Universe starting
from the radiation era to $\Lambda$CDM model. Again figure 7
represents the variation of $\Lambda$ against the scale factor
$a(t)$ with $\gamma_{1}=1, 3, 3.5$ and figure 8 represents the
variation $G$ against the scale factor $a(t)$. These figures show
that for this particular phenomenological model of $\Lambda$, $G$
starting from very low initial value increases largely and
becomes constant after a certain period of time, whereas
$\Lambda$ starting from a very large decreases largely to
reach a very low value and becomes constant.\\

\begin{figure}
\includegraphics[height=1.7in]{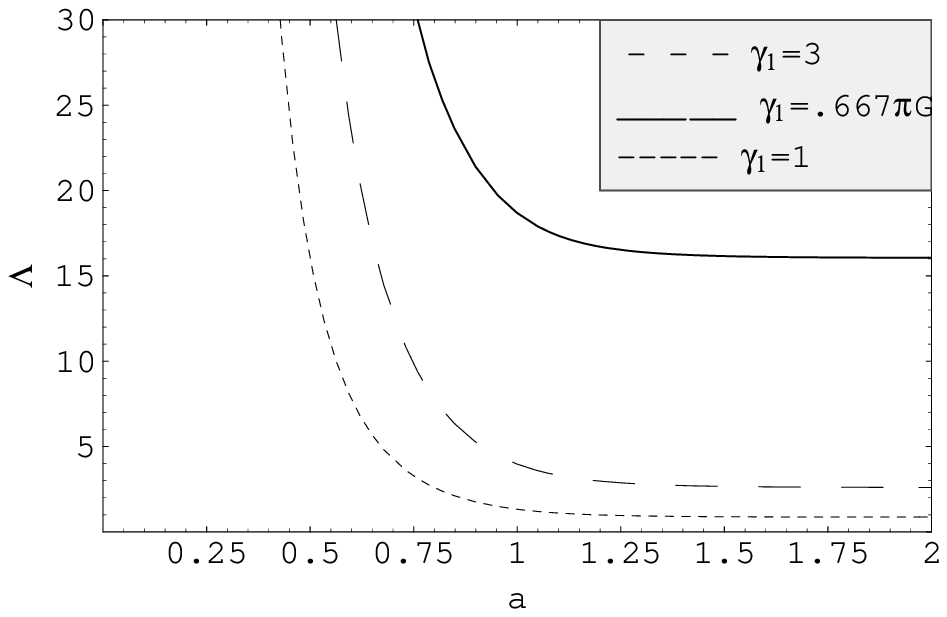}
\includegraphics[height=1.7in]{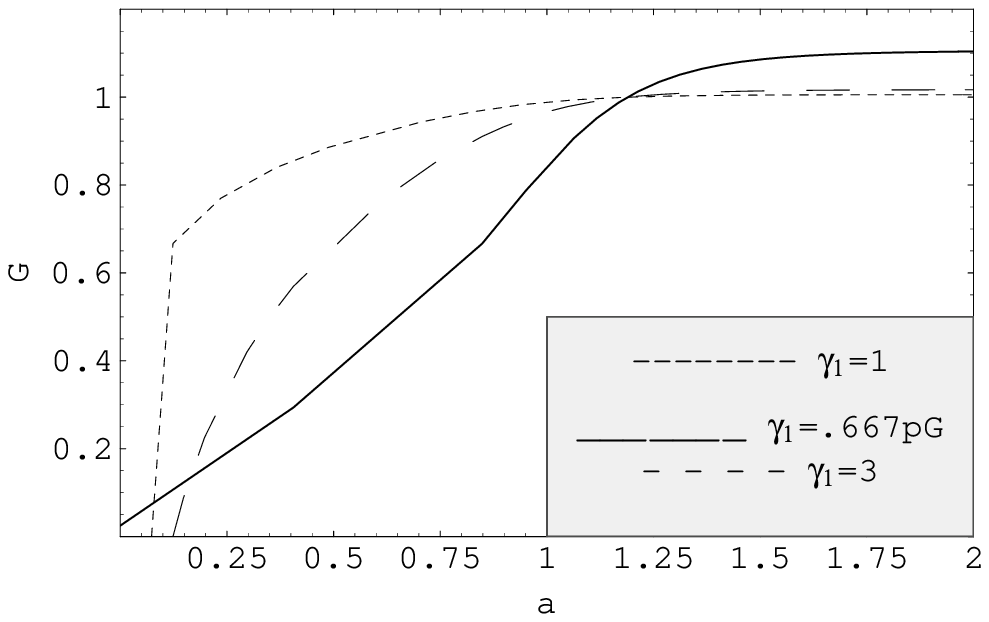}\\
\vspace{1mm}
Fig.7~~~~~~~~~~~~~~~~~~~~~~~~~~~~~~~~~~~~~~~~~~~~~~~~~~~~~Fig.8\\
\vspace{5mm}

\vspace{5mm} Fig. 7 shows the variation of $\Lambda$
 against $a(t)$ for different values of $\gamma_{1}=1, 3, 3.5$ and for
 $\alpha$ = ~1, $A=1/3$, $C_{1}=1$. Fig. 8 shows the variation of $G$ against $a(t)$ for
 different values of $\gamma_{1}=1, 3, 3.5$ and for
 $\alpha$ = ~1, $A~=~1/3$, $C_{1}~=~1, B~=~1,~C~=~1$. \hspace{14cm} \vspace{4mm}

\end{figure}

\subsection{\normalsize\bf{Model with $\Lambda\propto H^{2}$}}
We consider
\begin{equation} \Lambda=\gamma_{2} H^{2}
\end{equation}
Proceeding as above we get, \begin{equation} \Lambda=8 \pi G
\frac{\gamma_{2}}{3-\gamma_{2}}\rho  \end{equation} where $
\gamma_{2}$ is a constant.\\

Solving equation (22), (23) and (29) we get,
\begin{equation} G=\frac{C_{2}}{\rho^{\frac{\gamma_{2}}{3}}}
\end{equation}
where $C_{2}$ is a constant.\\

\begin{figure}
\includegraphics[height=1.7in]{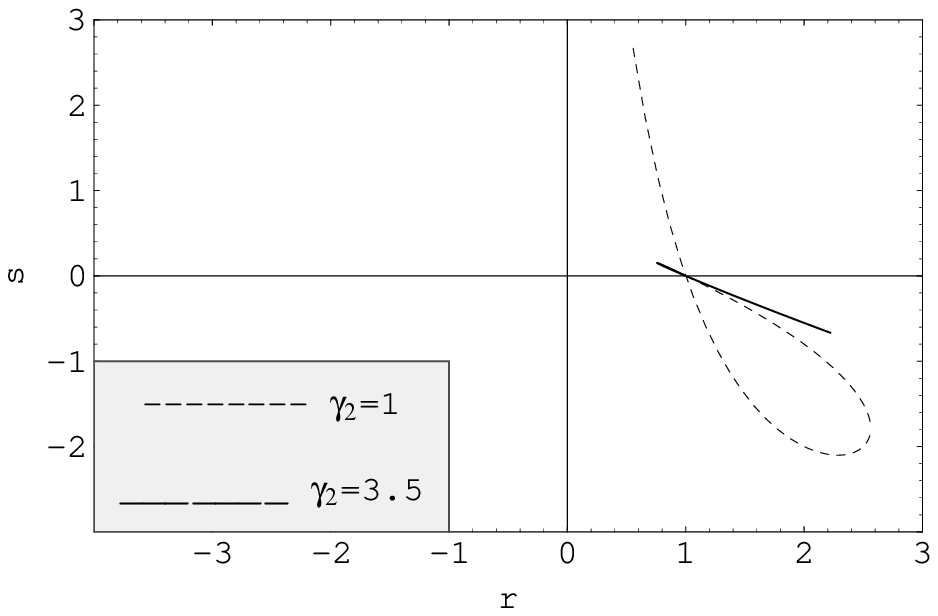}

Fig.9 \vspace{5mm}

\vspace{5mm} Fig. 9  shows the variation of $s$
 against $r$ for different values of $\gamma_{2}=1$ and $3.5$ and
$A=\frac{1}{3}, \alpha=1, B=1, C_{2}=1,~C~=~1$. \hspace{14cm}
\vspace{4mm}

\end{figure}

Using equations (2), (4) and (23), we find the state-finder
parameters as,
\begin{equation}
r=1+\frac{(1+y)(3-\gamma_{2})[3\{A(1+\alpha)-y\alpha\}-(1+y)\gamma_{2}]}{2}
\end{equation}
\begin{equation}
s=\frac{(1+y)(3-\gamma_{2})[3\{A(1+\alpha)-y\alpha\}-(1+y)\gamma_{2}]}{(3-\gamma_{2})y-\gamma_{2}}
\end{equation}
where $y=\frac{p}{\rho}$.\\

Now $q=\frac{1}{2}[(3-\gamma_{2})y-(\gamma_{2}-1)]]$. These
equations can further be resolved into a single relation of $r$
and $s$, which can be plotted diagrammatically in the ${r,s}$
plane. Here we get a discontinuity at $\gamma_{2}=3$. We have
plotted these values in the ${r,s}$ plane taking $\gamma_{2}=1$
and $3.5$ in figure 9 ($A=\frac{1}{3}, \alpha=1$). This case
explains the present acceleration of the Universe, starting from
radiation era to $\Lambda$CDM model.\\

\begin{figure}
\includegraphics[height=1.7in]{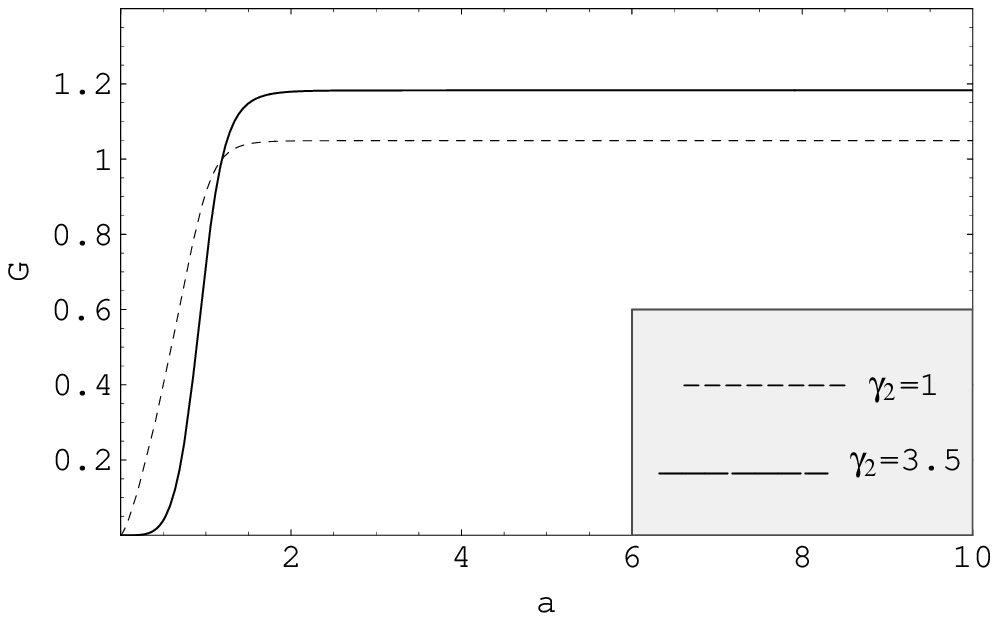}
\includegraphics[height=1.7in]{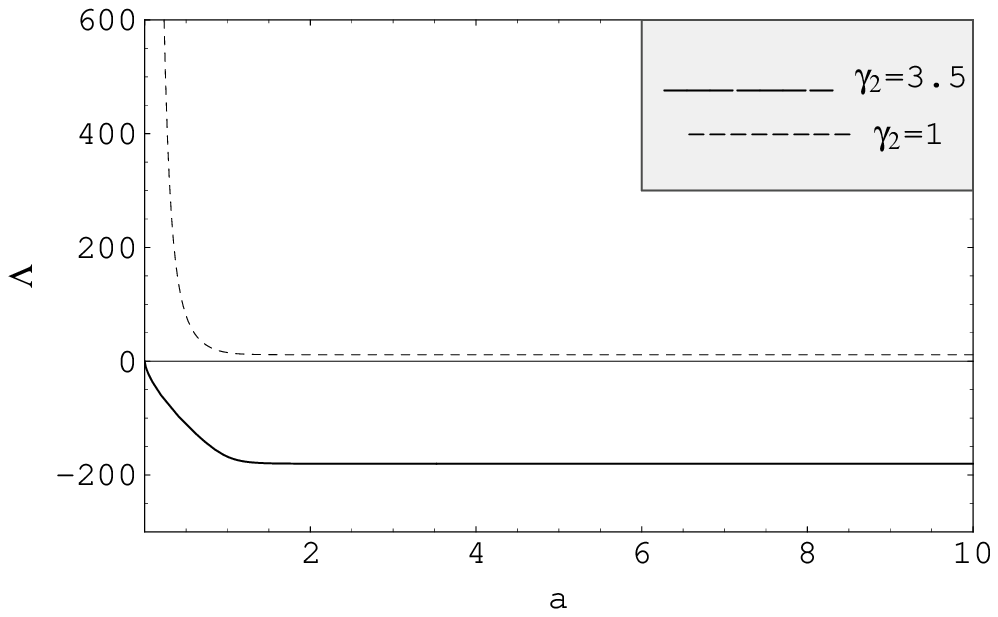}\\
\vspace{1mm}
Fig.10~~~~~~~~~~~~~~~~~~~~~~~~~~~~~~~~~~~~~~~~~~~~~~~~~~~~~Fig.11\\
\vspace{5mm}

\vspace{5mm} Fig. 10 shows the variation of $G$
 against $a(t)$ for different values of $\gamma_{1}=1, 3.5$ and for
 $\alpha$ = ~1, $A=1/3$, $C_{1}=1,~C~=1,~B~=~1$. Fig. 8 shows the variation of $\Lambda$ against $a(t)$ for
 different values of $\gamma_{1}=1, 3, 3.5$ and for
 $\alpha$ = ~1, $A~=~1/3$, $C_{2}~=~1, B~=~1,~C~=~1$. \hspace{14cm} \vspace{4mm}

\end{figure}

Also figures 10 and 11 show respectively the variation of $G$ and
$\Lambda$ against the scale factor for the same values of the
constants. Here also like the previous case $G$ starting from a
very low initial value increases largely and then continues to be
constant near unity. On the other hand $\Lambda$ starting from a
large value decreases largely and continues to be constant after
a certain period of time.\\

\subsection{\normalsize\bf{Model with $\Lambda\propto \frac{\ddot{a}}{a}$}}

Here we consider
\begin{equation}
\Lambda=\gamma_{3} \frac{\ddot{a}}{a}
\end{equation}
where $\gamma_{3}$ is a constant.\\

Using equation (33) in equations (4) and (5), we get,
\begin{equation}
\Lambda=-\frac{4 \pi G \gamma_{3}}{3-\gamma_{3}}(\rho+3p)
\end{equation}
Also, $G$ can be solved to be,
\begin{equation}
G=C_{3}[\rho^{\frac{1+3A}{2-\gamma_{3}(1+A)}}\{2-\gamma_{3}(1+A)+\frac{B
\gamma_{3}}{\rho^{\alpha+1}}\}^{\{-\frac{3\alpha}{\gamma_{3}(1+\alpha)}+\frac{1+3A}{(1+\alpha)(2-\gamma_{3}(1+A))}\}}]^{\frac{\gamma_{3}}{3}}
\end{equation}
Using equations (2), (4), (22), (23), we find,
\begin{equation}
r=1+\frac{(1+y)(3-\gamma_{3})[6\{A(1+\alpha)-y\alpha\}+\gamma_{3}(1+y)]}{[2-\gamma_{3}(1+y)]^{2}}
\end{equation}
\begin{equation}
s=\frac{2(1+y)(3-\gamma_{3})[6\{A(1+\alpha)-y\alpha\}+\gamma_{3}(1+y)]}{3[2-\gamma_{3}(1+y)][\gamma_{3}+(\gamma_{3}+6)y}
\end{equation}
where $y=\frac{p}{\rho}$ and $C_{3}$ is a constant. Equations
(36) and (37) can further be resolved to get one single relation
between $r$ and $s$ and plotted diagrammatically taking
$\gamma_{3}=2$ and $3.5$ (figure 12). Since deceleration parameter
$q=-\frac{\ddot{a}}{a
H^{2}}=-\frac{\lambda}{\gamma_{3}H^{2}}=\frac{4\pi
G}{(3-\gamma_{3}) H^{2}}$, is negative in the present epoch, we
get $3-\gamma_{3}<0$, i.e., $\gamma_{3}>3$. Also for
$\gamma_{3}=3$ we get discontinuity. Both the models represent
the phases of the Universe starting from radiation era to
$\Lambda$CDM model. Again $G$ and $\Lambda$ can be plotted against
$a$ (figures 13 and 14 respectively) . Unlike the previous cases
this model an opposite nature of $G$ and $\Lambda$, as $G$
decreases with time and $\Lambda$ increases with time.\\
\begin{figure}
\includegraphics[height=1.7in]{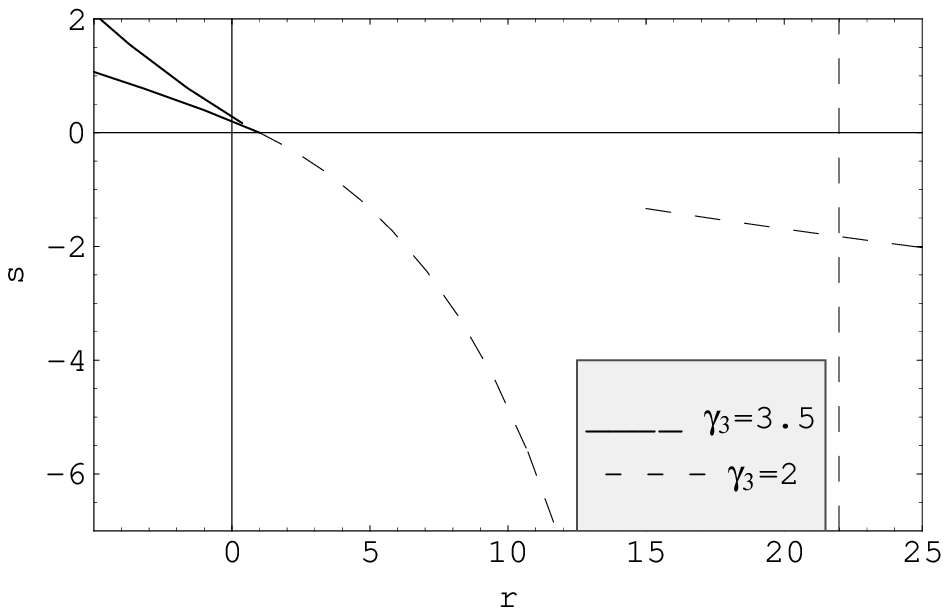}

Fig.12 \vspace{5mm}

\vspace{5mm} Fig. 12 shows the variation of $s$
 against $r$ for different values of $\gamma_{2}=2$ and $3.5$ and
$A=\frac{1}{3}, \alpha=1, B=1, C_{3}=1,~C~=~1$. \hspace{14cm}
\vspace{4mm}

\end{figure}

\begin{figure}
\includegraphics[height=1.7in]{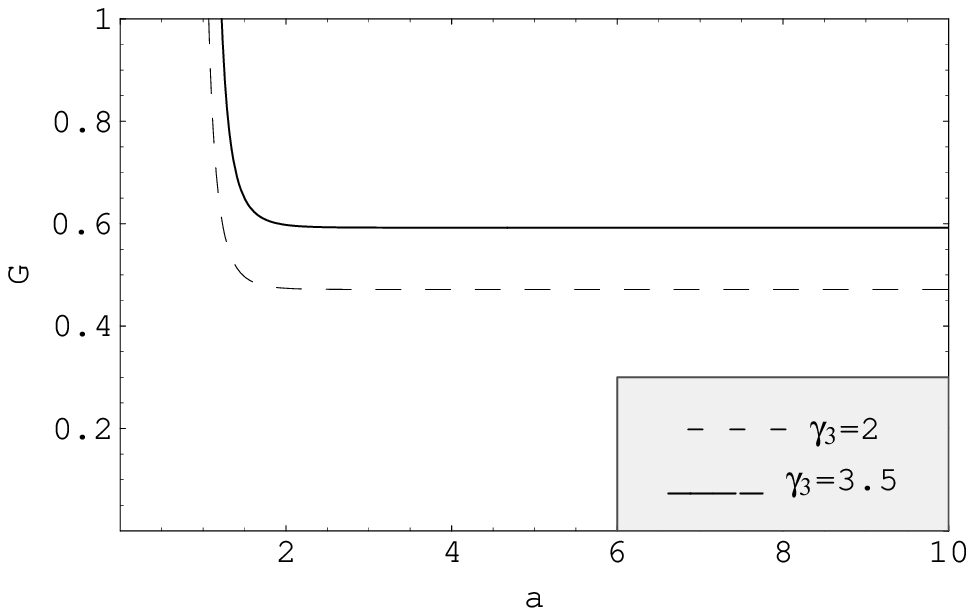}
\includegraphics[height=1.7in]{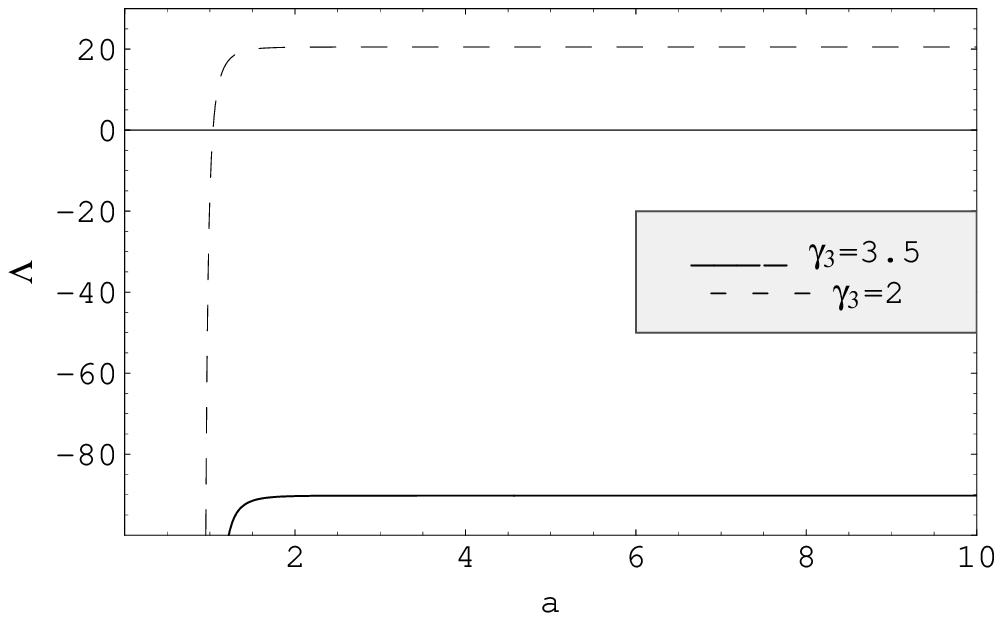}\\
\vspace{1mm}
Fig.13~~~~~~~~~~~~~~~~~~~~~~~~~~~~~~~~~~~~~~~~~~~~~~~~~~~~~Fig.14\\
\vspace{5mm}

\vspace{5mm} Fig. 13 shows the variation of $G$
 against $a(t)$ for different values of $\gamma_{1}=2, 3.5$ and for
 $\alpha$ = ~1, $A=1/3$, $C_{3}=1,~C~=1,~B~=~1$. Fig. 14 shows the variation of $\Lambda$ against $a(t)$ for
 different values of $\gamma_{1}=2, 3, 3.5$ and for
 $\alpha$ = ~1, $A~=~1/3$, $C_{3}~=~1, B~=~1,~C~=~1$. \hspace{14cm} \vspace{4mm}

\end{figure}

\section{\normalsize\bf{Discussion}}
Here we have considered three phenomenological models of
$\Lambda$, with or without keeping $G$ to be constant. Keeping
$G$ constant we always get accelerated expansion of the Universe.
For the first case, i.e., $\Lambda\propto \rho$ or more precisely,
$\Lambda=\beta_{1}\rho$, for particular choices of the constants
we get that the dark energy responsible for the present
acceleration is nothing but $\Lambda$. Also the density parameter
of the Universe for this case is given by,
$\Omega_{m}^{\beta_{1}}=\frac{8\pi G \rho}{3H^{2}}=\frac{8\pi
G}{8 \pi G +\beta_{1}}$ and the vacuum density parameter is
$\Omega_{\Lambda}^{\beta_{1}}=\frac{\Lambda}{3H^{2}}=\frac{\beta_{1}}{8\pi
G+\beta_{1}}$, so that
$\Omega_{total}=\Omega_{m}+\Omega_{\Lambda}=\Omega_{m}^{\beta_{1}}+\Omega_{\Lambda}^{\beta_{1}}=1$.
Also for $\Lambda\propto H^{2}$, i.e., $\Lambda=\beta_{2}~H^{2}$,
the density parameter and vacuum density parameter are given by,
$\Omega_{m}^{\beta_{2}}=\frac{3-\beta_{2}}{3}$ and
$\Omega_{\Lambda}^{\beta_{1}}=\frac {\beta_{2}}{3}$ respectively,
so that
$\Omega_{total}=\Omega_{m}^{\beta_{2}}+\Omega_{\Lambda}^{\beta_{1}}=1$.
Again for $\Lambda\propto \frac{\ddot{a}}{a}$ or
$\Lambda=\beta_{3}~\frac{\ddot{a}}{a}$, we have the corresponding
parameters as,
$\Omega_{m}^{\beta_{3}}=\frac{2(3-\beta_{3})}{3(2-\beta_{3}-\beta_{3}\frac{p}{\rho})}$,
$\Omega_{\Lambda}^{\beta_{3}}=\frac{-\beta_{3}(1+3\frac{p}{\rho})}{3(2-\beta_{3}-\beta_{3}\frac{p}{\rho})}$
and $\Omega_{total}=1$. Now
$\Omega_{total}=\Omega_{m}^{\beta_{3}}+\Omega_{\Lambda}^{\beta_{3}}=1$
for all the models. Also we can compare these models by taking,
$\Omega_{m}^{\beta_{1}}=\Omega_{m}^{\beta_{2}}$, so that
$\beta_{2}=\frac{3\beta_{1}}{(8\pi G +\beta_{1})}$. Now we would
like to take into account the present values of the density
parameter and vacuum parameter obtained by the recent
measurements. Considering $\Omega_{m 0}=0.33\pm.035$, we
calculate the present values of the proportional constants to be
$1.7397K \leq {\beta_{1}}^{0}\leq 2.3898 K,~1.905\leq
{\beta_{2}}^{0}\leq 2.115 $ and $3.7937 \leq {\beta_{3}}^{0}\leq
4.2099$, where $K= 8\pi G_{0}$ and $G_{0}$ is the present value of
the gravitational constant. Thus we get the value of
${\beta_{3}}^{0}$ to be lesser than the previous works [5, 14].
Again considering $G$ to be time-dependent, we get the same
values of the parameters as that with $G$ constant, i.e., the
ranges of ${\gamma_{1}}^{0}, {\gamma_{2}}^{0}, {\gamma_{3}}^{0}$
are same as that of ${\beta_{1}}^{0}, {\beta_{2}}^{0},
{\beta_{3}}^{0}$ respectively. Here also we get cosmic
acceleration and the nature of variation $G$ and $\Lambda$ as
well. We get two different cases regarding the variation of $G$
and $\Lambda$. For the first two cases we see that $G$ increases
and $\Lambda$ decreases with time, whereas for the third case $G$
decreases and $\Lambda$ increases with time. In all the cases the
values become constant after a certain period of time, i.e.,the
present day values of $G$ and $\Lambda$ are constants. Thus these
models with the phenomenological laws give us some interesting
features of the cosmic acceleration and some modified values of
the parameters. Also we get the natures of the Cosmological
Constant and the Gravitational Constant over the total age of the
Universe. We can also make use of the statefinder parameters to
show the evolution of the Universe starting from radiation era to
$\Lambda$CDM model.\\

{\bf References:}\\
\\
$[1]$ P. S. Wesson (1978), {\it Cosmology and Geophysics} (Oxford
: Oxford University Press ); P. S. Wesson (1980), {\it Gravity,
Particles and Astrophysics} ( Dordrecht : Rieded ).\\
$[2]$ P. A. M.  Dirac, {\it Proc. R. Soc. A} {\bf 165} 119 (1938)
; {\bf 365} 19 (1979) ; {\bf 333} 403 (1973); The General Theory
of Relativity (New York : Wiley) 1975.\\
$[3]$ S. Weinberg, {\it Rev. Mod. Phys.} {\bf 61} 1 (1989); S. M.
Carroll, W. H. Press and E. L. Turner , {\it Ann. Rev. Astron.
Astrophys.} {\bf 30} 499 (1992).\\
$[4]$ K. Freese, F. C. Adams, J. A. Freeman and E. Mottola, {\it
Nucl. phys. B.} {\bf 287} 797 (1987); M. Ozer and M. O. Taha,
{\it Nucl. phys. B.} {\bf 287} 776 (1987); M. Gasperini, {\it
Phys. Lett. B} {\bf 194} 347 (1987); {\it Class. Quantum Grav.}
{\bf 5} 521 (1998); W. Chen and Y. S. Wu, {\it Phys. Rev. D} {\bf
41} 695 (1990).\\
$[5]$ A. Banerjee, S. B. Dutta Chaudhuri and N. Banerjee, {\it
Phys. Rev. D} {\bf 32} 3096 (1985); O. Bertolami {\it Nuovo
Cimento} {\bf 93B} 36 (1986); {\it Fortschr. Phys.} {\bf 34} 829
(1986); Abdussattar and R. G. Vishwakarma, {\it Class. Quantum
Grav.} {\bf 14} 945 (1997); Arbab I. Arbab, {\it Class. Quantum Grav.} {\bf 20} 93 (2003); D. Kalligas, P. Wesson and C. W. F. Everitt, {\it Gen. Rel. Grav.} {\bf 24} 351 (1992).\\
$[6]$ A. S. Al-Rawaf and M. O. Taha, {\it Gen. Rel. Grav.} {\bf
28} 935 (1996).\\
$[7]$ N. A. Bachall, J. P. Ostriker, S. Perlmutter and P. J.
Steinhardt, {\it Science} {\bf 284} 1481 (1999).\\
$[8]$ S. J. Perlmutter et al, {\it Astrophys. J.} {\bf 517} 565
(1999).\\
$[9]$ A. Kamenshchik, U. Moschella and V. Pasquier, {\it Phys.
Lett. B} {\bf 511} 265 (2001); V. Gorini, A. Kamenshchik, U. Moschella and V. Pasquier, {\it gr-qc/0403062}.\\
$[10]$ V. Gorini, A. Kamenshchik and U. Moschella, {\it Phys. Rev.
D} {\bf 67} 063509 (2003); U. Alam, V. Sahni , T. D. Saini and
A.A. Starobinsky, {\it Mon. Not. Roy. Astron. Soc.} {\bf 344},
1057 (2003); M. C. Bento, O. Bertolami and A. A. Sen, {\it Phys.
Rev. D} {66} 043507 (2002).\\
$[11]$ H. B. Benaoum, {\it hep-th}/0205140.\\
$[12]$ U. Debnath, A. Banerjee and S. Chakraborty, {\it Class.
Quantum Grav.} {\bf 21} 5609 (2004).\\
$[13]$ V. Sahni, T. D. Saini, A. A. Starobinsky and U. Alam, {\it
JETP Lett.} {\bf 77} 201 (2003).\\
$[14]$ U. Mukhopadhyay, S. Ray, astro-ph/0407295.\\

\end{document}